\RequirePackage{fix-cm}
\documentclass[twocolumn]{svjour3}          
\usepackage{graphicx}
\usepackage{mathrsfs}
\usepackage{color}

\smartqed  
\usepackage{graphicx}
\usepackage{mathrsfs}
\usepackage{color}

\newcommand{\ka}{\mathbf{k}}

\journalname{myjournal}

\begin{document}

\title{Bloch oscillations: Inverse problem
}
\author{Jos\'e~A. Gonz\'alez$^1$         \and
        Sa\'ul Hern\'andez-Ortiz$^1$     \and
        Carlos~E.~L\'opez$^1$             \and
        Alfredo Raya$^1$
}


\institute{Jos\'e~A. Gonz\'alez \at
              gonzalez@ifm.umich.mx           
           \and
           Sa\'ul Hern\'andez-Ortiz \at
              sortiz@ifm.umich.mx
           \and
           Carlos~E.~L\'opez \at
              clopez@ifm.umich.mx 
           \and
           Alfredo Raya \at
              raya@ifm.umich.mx
          \and
            $^1$Instituto de F\'isica y Matem\'aticas, Universidad Michoacana de San Nicol\'as de Hidalgo.
}

\date{Received: date / Accepted: date}

\maketitle

\begin{abstract}
We use a neural network approach to explore the inverse problem of Bloch oscillations in a mo\-no\-a\-to\-mic linear chain: given a signal describing the path of oscillations of electrons as a function of time,  we determine the strength of the applied field along the direction of motion or, equivalently, the  lattice spacing. We find that the proposed approach has more than $80\%$ of accuracy classifying the studied physical parameters.
\end{abstract}

\section{Introduction}

Understanding the motion of Bloch electrons in crystal structures is one of the traditional subjects of solid state physics, for which a number of techniques have been developed. The wave-particle duality of electrons in a crystal plays an essential role. Particularly, in the presence of an electric field, electrons are accelerated and its wavelength is shortened in such a manner that when it is twice the lattice spacing, it undergoes Bragg diffraction in reciprocal and real spaces, hence describing a periodic motion whose period depends entirely on the periodicity of the crystal along the field direction and the strength of the applied field itself~\cite{bloch,zener}. This periodic motion  has been dubbed Bloch oscillations (BO) and its description is known for almost one century from the seminal work of Bloch~\cite{bloch}. In real solids, intraband tunneling and ultrafast electron scattering prevent BO from being observed. Nevertheless, in high-purity semiconductor superlattices this phenomenon has been directly observed under a variety of  experimental conditions~\cite{exp1,exp2,exp21,exp22,exp3,exp31,exp4}.  The smoking gun is a plot of a certain displacement of electron position with sinusoidal profile in time. 

Moreover, within thefield of quantum simulators, where the main idea is that quantum dynamics of a givensystem can be reproduced by an equivalent optical system which can be quantum or classical. BO can be observed in a variety of different equivalent systems to bulk crystals, like atomic systems \cite{Dahan,Genske}, dielectric \cite{Pertsch,Morandotti,Sapienza} and plasmonic \cite{Block}waveguidesarrays and so on. Therefore, in this article we exploit the mathematical structure of the equations governing BO which can extended to other systems

In this article, we discuss the {\em inverse} BO problem (IBO): Given an {\em ``experimental''},  periodic curve representing the real space motion of electrons in a solid subject to an external electric field as a function of time and assuming that BO drive the dynamics, we determine the strength of the applied field or, equivalently, the lattice spacing at fixed electric field intensity. We use the linear chain as our working example and consider the nearest (1N) and next-to-nearest (2N) neighbors tight binding approximation.

 Although a plethora of methods to analyze periodic curves are known and well established, in this article we  construct an efficient machine learning  algorithm based on Artificial Neural Networks (ANN)~\cite{ANN,dsp}. The basic idea of the ANNs consists in simulate the complex structure of the biological neurons with a simplified  interconnection of computational nodes whose main objective is to process a set of signals and learn to classify them. When this is done in a proper way, the ANN can produce accurate predictions for signals that have never been presented to the network. This method of classification can be very efficient and accurate.
 
To address the issue of IBO, the remaining of the article is organized as follows: Section 2 presents the semiclassical framework to describe BO in a monoatomic lineal chain. Section 3 is devoted to introduce the setup of our ANN. Section 4 presents the data processing analysis and  final remarks are given in Section 5.

\section{Bloch oscillation: Semiclassical approach}
We start our discussion from the tight-binding Hamiltonian of a monoatomic lineal chain of lattice spacing $a$. At the 1N approximation, we have
\begin{eqnarray}
H \psi_n(\ka) &=& - t \psi_{n+1}(\ka) - t \psi_{n-1}(\ka) + \epsilon_0 \psi_{n}(\ka)\nonumber\\
&\equiv&\mathcal{E}^{(n)}(\ka) \psi_n(\ka)\;,
\end{eqnarray}
where $t$ is the hopping parameter. From Bloch theorem, it is straightforward to find that in this case, the energy-momentum dispersion relation is
\begin{equation}
    \mathcal{E}^{(n)}(\ka) = \epsilon_0  - \epsilon^{(n)}(\ka)\;, 
\end{equation}
where
\begin{equation}
 \epsilon^{(n)}(\ka) =  w (1 - \cos(|\ka|a))\;, 
\label{en}
\end{equation}
$\epsilon_0$ is the on-site energy and $w=2t$. Similarly, for the 2N neighbors approximation,
\begin{eqnarray}
    H \psi_n(\ka) &=& - t_2 \psi_{n+2}(\ka) - t_2 \psi_{n-2}(\ka)\nonumber\\
&& - t_1 \psi_{n+1}(\ka) - t_1 \psi_{n-1}(\ka) + \epsilon_0 \psi_{n}(\ka)\nonumber\\
&\equiv& \mathcal{E}^{(nn)}(\ka)\psi_n(\ka)\;.
\end{eqnarray}
In this case, $t_2$ and $t_1$ are the hopping for the corresponding neighbor. Then, from Bloch theorem we have that the dispersion relation is
\begin{equation}
\mathcal{E}^{(nn)}(\ka) = \epsilon_0  - \epsilon^{(nn)}(\ka)\;, 
\end{equation}
but in this case
\begin{equation}
\epsilon^{(nn)}(\ka) =  w (1 - \cos(|\ka|a)- w^\prime \cos(2|\ka|a))\;,
\label{enn}
\end{equation}
and $ w = 2t_1$ while $w^\prime = t_2/t_1.$

Next, we recall the semiclassical equations of motion  for an electron moving in an external electric field $\mathbf{E}$ oriented parallel to the linear chain, 
\begin{eqnarray}
 \frac{d\mathbf{k}}{dt} & = & -e\mathbf{E}\;, \label{em1}\\
 \frac{d\mathbf{r}}{dt} & = & \frac{1}{\hbar}\frac{\partial}{\partial\mathbf{k}}\epsilon(\ka)\;,\label{em2}
\end{eqnarray}
where in place of $\epsilon(\ka)$ we can insert $\epsilon^{(n)}(\ka)$ or $\epsilon^{(nn)}(\ka)$ from Eqs.~(\ref{en}) or (\ref{enn}), respectively.
We can straightforwardly integrate the equations of motion and obtain the velocities and trajectories for a given external field strength. 
Considering the chain oriented along the $x$ axis and a uniform electric field $\mathbf{E}=E\hat{e}_x$,  we integrate Eq. (\ref{em1}) assuming the initial condition $k(0) = 0$. Thus
\begin{equation}
        k(t)=-\frac{eE}{\hbar}t,
\end{equation}
where $k$ is the crystal-momentum of electrons in the linear chain.
On the other hand, the electron velocity is given in Eq.~(\ref{em2}), which  for 1N is
\begin{eqnarray}
        v_{(n)}(k(t)) & =& \frac{w a}{\hbar}\sin(k(t) a),\nonumber\\
          &=& -\frac{w a}{\hbar}\sin\left(\frac{eEa}{\hbar}t\right).\label{veln}
\end{eqnarray}
Finally, we obtain the profile of BO simply integrating  Eq. (\ref{veln}), which yields
\begin{eqnarray}
        x_{(n)}(t) & = & \frac{w}{eE}\cos\left(\frac{e E }{\hbar}a t\right),\nonumber \\
        & = & \frac{w}{eE}\cos(\omega_E t),
        \label{1v}
\end{eqnarray}
with $\omega_E =  e E a/ \hbar$. Analogously, for the 2N, we straightforwardly obtain
\begin{eqnarray}
        v_{(nn)}(t) & = & -\frac{w a}{\hbar}\left(\sin\left(\omega_E t\right)+2w^\prime\sin\left(2\omega_E t\right)\right), \label{vnn}\\
        x_{(nn)}(t) & = & \frac{w}{eE}\left(\cos\left(\omega_E t\right)+ 2 w^\prime\cos^2\left(\omega_E t\right)\right).
        \label{2v}
\end{eqnarray}
Equations~(\ref{1v}) and~(\ref{2v}) give a fairly good agreement to the experimental observations of BO \cite{exp1,exp2,exp21,exp22,exp3,exp31,exp4}.
In the next section we introduce artificial neural networks as the tool to analyze these oscillations.

\section{Artificial Neural Networks}
Among the different machine learning algorithms available nowadays, the artificial neural networks algorithm is a very useful and common option to solve classification problems. In this section we describe the implementation we use to classify  BO. Then we set the problem of IBO to test the performance of the approach. For simplicity, let us describe the method when we consider only the interaction of the 1N. The parameter $a$ is fixed and the magnitude of the electric field $E$ changes from 0.01 to 1 in steps of $\Delta E=0.01$, i.e. $E^p=p\cdot\Delta E$ with $p=1,...,P$ and $P=100$, in units where $\hbar=1=e$.

We choose a discrete set of points in time $t_i$ and a given value of the magnitude of the electric field $E^p$. Evaluating Eq. (\ref{1v}) with $a$ and $t_i$ given for every $E^p$ produces a set of discrete values for the position $x^p_i$. The upper index $p$ reflects the dependence of the position of the electron on the electric field. The pairs $(E^p, x^p(t_i)=x^p_i)$ are called the training and validation sets. From the total number of generated trajectories, 70\% of them are in the training set and the other 30\% are in the validation set.

Now, we use an ANN to classify the BO in terms of the electric field intensity. As a first approach, we introduce the values $x^p_i$ as inputs of the ANN and propagate them forward obtaining as a result a value that we interpret as the corresponding electric field $\tilde{E}^p$. A more detailed analysis shows that in order to improve the efficiency and accuracy of the network, instead of using all the values $x^p_i$ as inputs, it is possible to compute a discrete fast Fourier transform (DFFT) of the trajectory and only use the maximum value of the amplitude and the corresponding frequency. This approach reduces the number of inputs to only two.

The structure of the ANN consists in three layers (input, hidden and output) of neurons. All the neurons of one layer are connected to neurons of the next layer with weights. Each of these weights is mathematical represented with a number, initialized randomly between -1 and 1. Every neuron has an activation function that it is used to decide how the information is propagated into the next layer. The activation functions used in this paper are logistic and linear.

After propagating forward (from the inputs to the outputs) the amplitude and frequency of the trajectory, the obtained result $\tilde{E}^p$ in general does not correspond to the expected electric field $E^p$ because the ANN is initialized with random coefficients. To achieve accuracy, we need to perform an optimization process where the coefficients of the network are updated in order to classify in a proper way all the elements of the training set. First, we define a cost function as
\begin{equation}
c= \frac{1}{2} \sum_{k=1}^{K}(E^p_k-\tilde{E}^p_k)^2 \, ,
\end{equation}
where $K$ is the number of output neurons and depends on the number of parameters predicted by the ANN.

We perform a minimization process using offline backpropagation algorithm with a learning parameter $\gamma$ \cite{ANN}. Given the trajectory of the BO for one value of $E$, the internal parameters of the ANN are adjusted such that the predicted value $\tilde{E}^p$ gets as close as possible to $E^p$. Afterward, we perform the same process over the remaining elements of the training set to achieve a good accuracy predicting all the possible values of the parameter $E$ in the sample. Next, we use a different set of trajectories, which we call validation set, to track the behavior of the training process. If the validation error starts growing or the training error reaches a given threshold, then we stop the training process and move on to the prediction phase. 
After the training process is complete, we use new signals to test the accuracy of the predictions of the ANN. This set has the same amount of signals than the validation set and is created using random values for $E$ in the given interval and it is called the prediction set. In this stage the IBO problem is posed.
We are ready now to use these structures to the IBO analysis. In the next section we present technical details. \\

\section{Data processing} 
The signals were discretized considering the sampling theorem \cite{dsp}. This theorem establishes that the sample rate has to be at least twice the maximum frequency of the signal. In our case,
\begin{equation}
    \omega_E=E a \, ,
\end{equation}
and so we use
\begin{equation}
    \Delta t =\frac{1}{2 E a} .
\end{equation}

\subsection{Bloch Oscillations for 1N interactions}
\subsubsection{Variable $E$ ($a$ fixed)}

\begin{figure}[h]
\centerline{
\includegraphics[width=0.45\columnwidth]{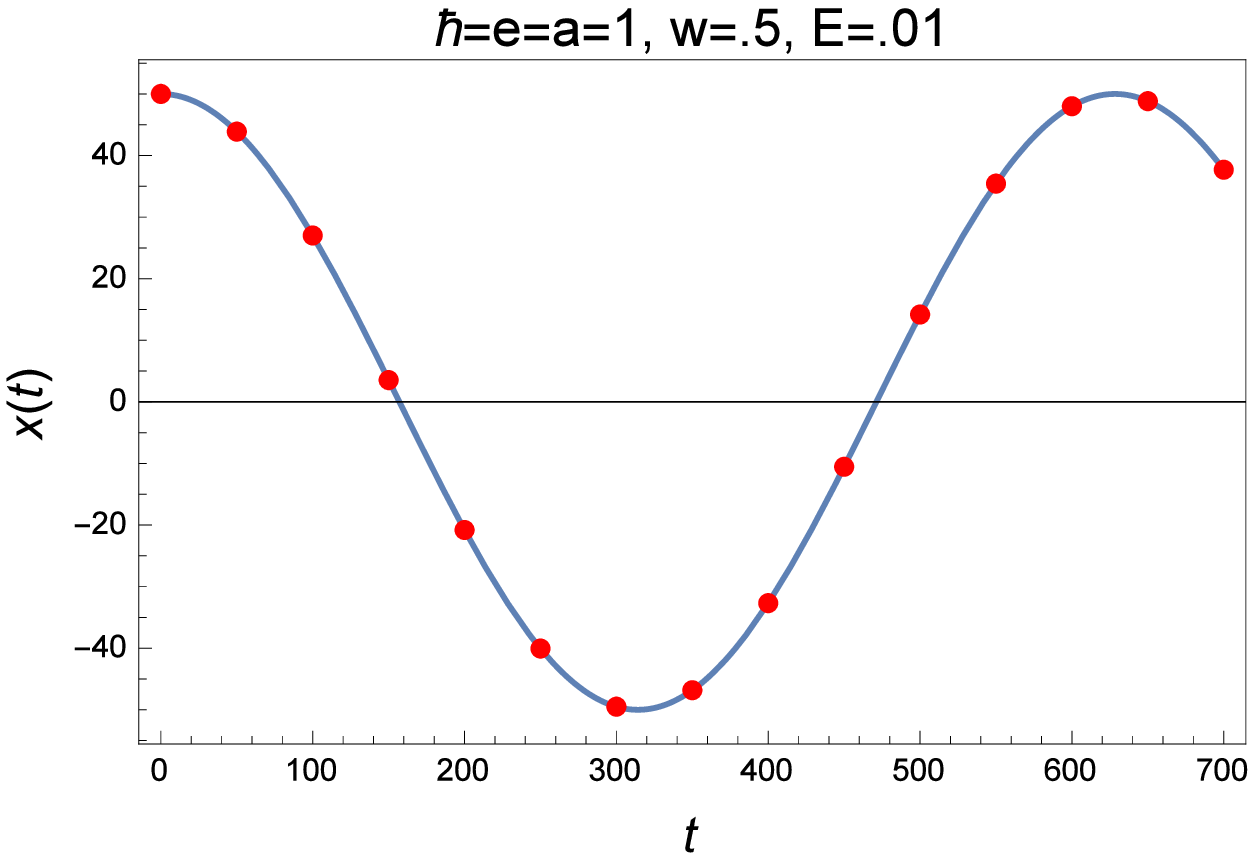}
\includegraphics[width=0.45\columnwidth]{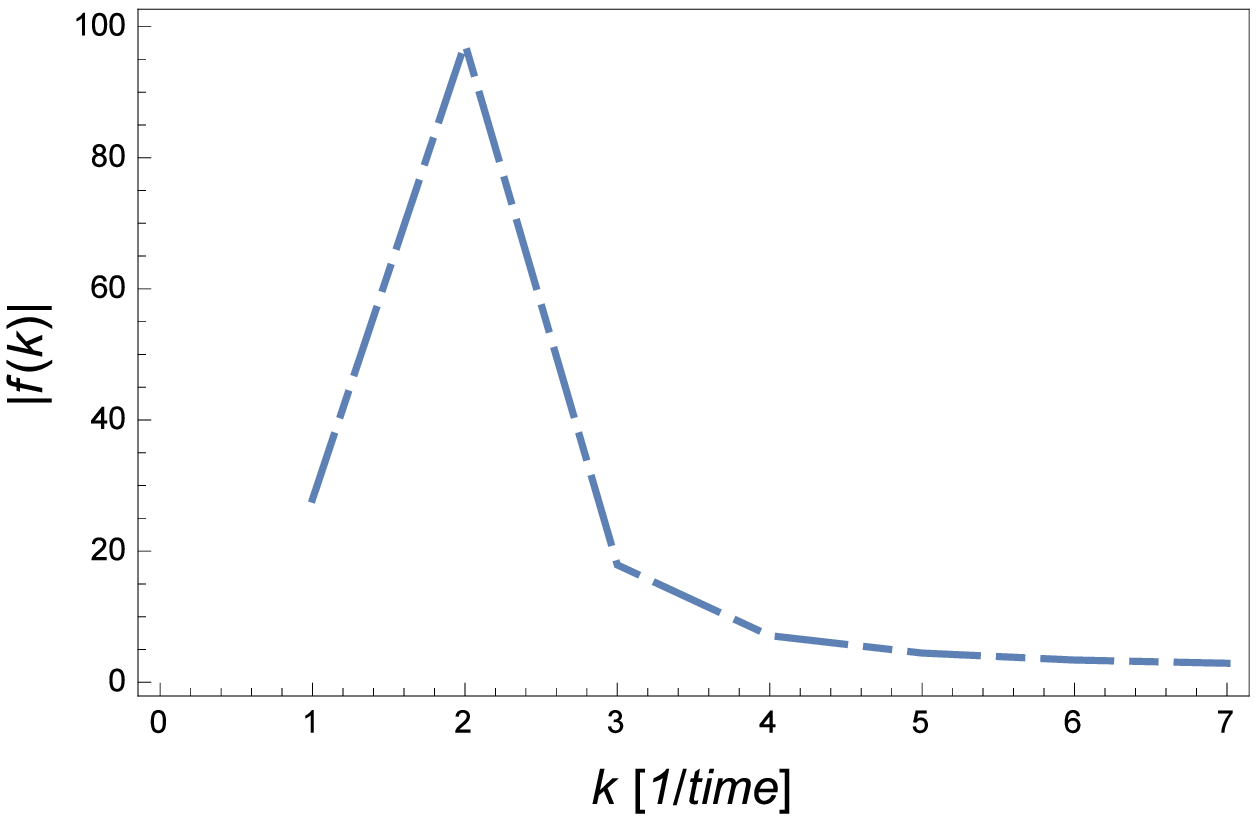}
}
\caption{{\em Left:} Trajectory for a 1N scenario corresponding to the lowest $\omega_E$ ($a=1.0$, $w=0.5$, $E=.01$). The dots show the positions where the DFFT is computed. {\em }Right: Absolute value of the DFFT. The selected frequency and amplitude are $k^1=2$ and $|f(2)|$.}
\label{osctofrec}
\end{figure}

One hundred signals of BO were generated using Eq.~(\ref{1v}), varying only $E$ ($0.01 \le E \le 1$) with steps of $\Delta E = 0.01$. All the remaining parameters were fixed ($a=1.0,w=0.5$). The total number of $t_i$ is equal to $700$ to ensure that the signal created with the lowest frequency, at least has one period (Figure \ref{osctofrec}). 

A DFFT $f(k)$ is computed over each trajectory and the maximum value of the amplitude and the corresponding frequency of the signal were used as inputs for the network. It is computed as
\begin{equation}
f(k) = \sum_{j=0}^{N} x^p(t_j) {\rm Exp}\left[\frac{-2\pi i k j}{N}\right] \, ,
\label{dfft}
\end{equation}
where the indexes $j$ and $k$ are related with the discrete values of time and frequency. In the case of the time, we change the index $i$ by $j$ to avoid confusions with the complex number $i$ appearing in Eq. (\ref{dfft}). The relation between $k$ and the frequency is given by $\nu_k=\frac{k}{N}2\omega_E$. We use a super-index $p$ to denote the energy associated with the signal.

The amplitude and frequency are rescaled before using them as inputs of the ANN using the following transformation
\begin{equation}
    \tilde{X}=\frac{X-\frac{1}{P}\sum X}{\rm{max}(X)-\rm{min}(X)} \, ,
    \label{eq:rescale}
\end{equation}
where the maximum and minimum values of $X$ are computed over all the elements on the training set. Substituting $k^p$ and $|f(k^p)|$ instead of $X$ gives us the two required inputs for the network.

The ANN classifies the input data between ten possible ranges for the electric field.
We call each of these ranges a class and they are parametrized by the index $cl=1,...,10$.  This means that the output generated by the network will be associated with a class, i.e., with a corresponding electric field strength. The output neuron uses a linear activation function, so that values outside the interval $[0,1]$ are associated with the first or last class. The value of the targets associated with each class is given by the relation
\begin{equation}
    T_1^p= 0.1 cl-0.05 \, ,
\end{equation}
where the numbers $0.1$ and $0.05$ come from the fact that we divide the output interval in 10 classes. The total number of steps used to train the network is $S=10,000$ and the learning parameter is chosen as $\gamma=0.0054$. 
The training process is repeated five times to verify a good performance and avoid  strange initializations of weights. The weights obtained from the training are used to test the performance for every set. 

To decide if the prediction is correct,  we consider the following rule. If the electric field corresponding to the configuration lies in the range
\begin{equation}
    T^p_1-0.05<E^p<T^p_1+0.05
    \label{tar1N}
\end{equation}
then the output is taken as a correct classification, which means that the value $E$ that generates the signal is in the range of values that corresponds to that class. Otherwise is consider a wrong classification.
Once the network is completely trained over the training set, we use a set of generated trajectories with random values of $E$ between [$0,1$] to verify the prediction capabilities of the ANN. The average over five different prediction sets and the efficiency of the ANN for all the sets is shown in Table \ref{1NE}.

\begin{table}[h]
\begin{center}
\begin{tabular}{|c|r@{.}l|r@{.}l|r@{.}l|}
\hline
&\multicolumn{2}{|c|}{Training set} & \multicolumn{2}{|c|}{Validation set}&\multicolumn{2}{|c|}{Test set} \\
\hline
CCP&\multicolumn{2}{|c|}{$70$}&\multicolumn{2}{|c|}{$30$}&\multicolumn{2}{|c|}{$29$}\\  
\hline
Percent &\multicolumn{2}{|c|}{$1$}&\multicolumn{2}{|c|}{$1$}&\hspace{.2cm}0&966\\
\hline
\end{tabular}
\caption{Correctly Classified Patterns (CCP) of the ANN for the signals generated varying $E$ between [$0,1$]. We present the results for 70 training patterns, 30 validation patterns and the average of 5 sets of prediction of 30 patterns each.}
\label{1NE}
\end{center}
\end{table}
\subsubsection{Variable a (E fixed).}
We repeat exactly the same procedure as described above, but now fixing the value of the electric field $E=1$ and allowing the parameter $a$ to change between $0.505$ and $1$. One hundred trajectories were generated using steps of $\Delta a=0.005$. 

The classes are selected with the condition
\begin{displaymath}
    \frac{cl-1}{2}<a-0.5\le \frac{cl}{20} \, .
\end{displaymath}
As before, the initial weights were generated randomly between [$-1,1$] and the total of steps for the training was $S=10000$ with a learning rate of $\gamma=0.0055$. The efficiency of the ANN for this scenario is presented on Table \ref{1Na}.

\begin{table}[h]
\begin{center}
\begin{tabular}{|c|r@{.}l|r@{.}l|r@{.}l|}
\hline
&\multicolumn{2}{|c|}{Training set} & \multicolumn{2}{|c|}{Validation set}&\multicolumn{2}{|c|}{Test set} \\
\hline
CCP&\multicolumn{2}{|c|}{$65$}&\multicolumn{2}{|c|}{$27$}&\hspace{.4cm}28&6\\ 
\hline
Percent &\hspace{.5cm}0&928&\hspace{.8cm}0&9&0&953\\
\hline
\end{tabular}
\caption{Correctly classified patterns (CCP) of the ANN for the signals generated varying $a$ between [$0.5,1$]. We present the results for 70 training patterns, 30 validation patterns and the average of 5 sets of prediction of 30 patterns each.}
\label{1Na}
\end{center}
\end{table}

\subsection{Bloch Oscillations for 2N interactions}
\begin{figure}[h]
\centerline{\includegraphics[width=0.45\columnwidth]{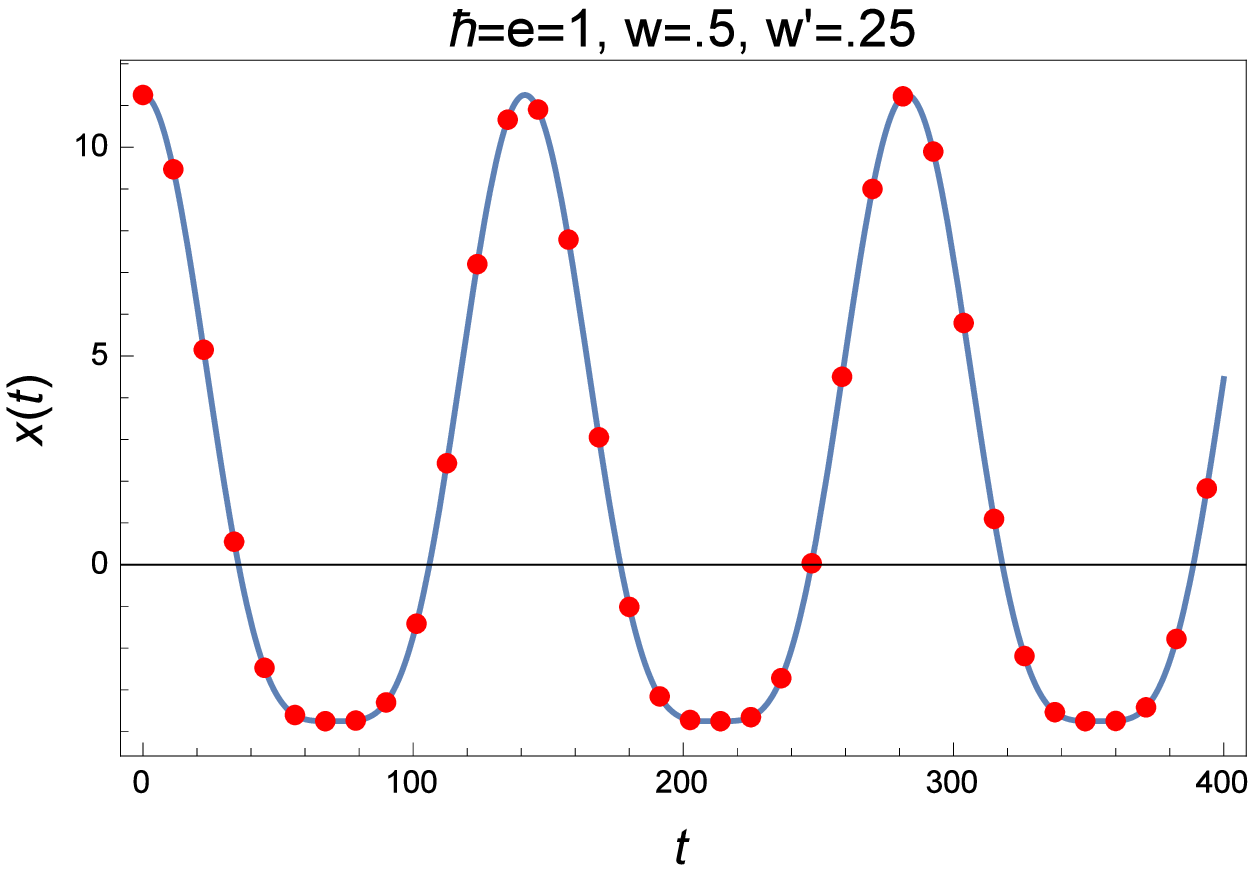}
\includegraphics[width=0.45\columnwidth]{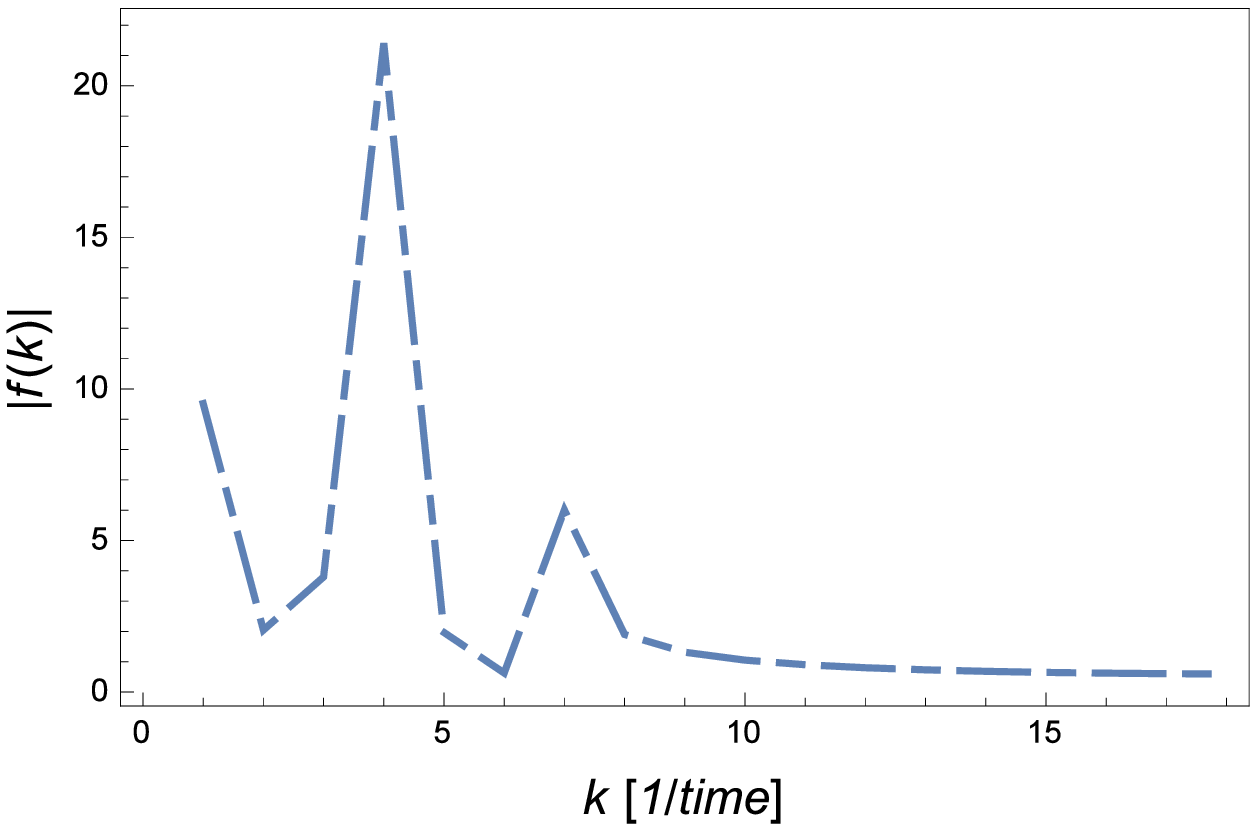}
}
\caption{{\em Left:} Trajectory for 2N generated using $w=0.5,w'=0.25$, $a=.66$, $E=0.066$ corresponding to $p=76$. The dots show the positions where the DFFT is computed. {\em Right:} Absolute value of the DFFT. The selected frequencies and amplitudes are $k^{76}_{1}=1$, $k^{76}_{2}=4$, $k^{76}_{3}=7$ and $|f(2)|$, $|f(4)|$, $|f(7)|$ respectively.}
\label{osctofrec2}
\end{figure}

When 2N interactions are taken into account, signals are generated using Eq. (\ref{2v}), varying simultaneously $a$ and $E$ with the remaining parameters fixed ($w=1/2,w'=1/4$). In these cases the network has two neurons in the output layer associated with the values of $a$ and $E$. The range of the parameters studied in this scenario is $0.5333 \le a \le 1$ with $\Delta a=0.0333$ and $0.0666 \le E \le 1$  with $\Delta E =0.0666$. Therefore, 225 signals were generated (Figure \ref{osctofrec2}) each of them with a time length of $400$. 

Six values are stored, three of these values corresponding to the frequencies ($k_{1}, k_{2}, k_{3}$) and the other three corresponding to their amplitudes ($|f(k_{1})|$, $|f(k_{2})|$, $|f(k_{3})|$). As before, Eq. (\ref{eq:rescale}) is used to rescale the values of the amplitudes and the frequencies.
Instead of using $10$ classes, now we use 3 for each parameter which means in total 9 classes, where the indexes are $cl_a$ for $a$ and $cl_E$ for $E$. The number of classes was selected in such a way that the training and prediction time of the ANN is similar to the previous case.
The corresponding targets are:
\begin{equation}
    T_1^p=\frac{1}{3}\bigg(cl_a-\frac{1}{2}\bigg)\,\,\, , \hspace{0.5cm}
    T_2^p=\frac{1}{3}\bigg(cl_E-\frac{1}{2}\bigg)\, .
\end{equation}

The parameters for the training phase are weights between [$-1,1$], $S=20000$, $\gamma=.00022$ and the conditions to decide if the classification is correct are
\begin{equation}
    T^p_1-\frac{1}{6}<O^p_1<T^p_1+\frac{1}{6}\,\,\, , \hspace{0.5cm}
    T^p_2-\frac{1}{6}<O^p_2<T^p_2+\frac{1}{6} \, .
\end{equation}

The efficiency of the network for this experiment is in Table \ref{2N}.

\begin{table}[h]
\begin{center}
\begin{tabular}{|c|r@{.}l|r@{.}l|r@{.}l|r@{.}l|r@{.}l|r@{.}l|}
\hline
&\multicolumn{4}{|c|}{Training set} & \multicolumn{4}{|c|}{Validation set}&\multicolumn{4}{|c|}{Test set} \\
\hline
Output&\multicolumn{2}{|c|}{$O_1$}&\multicolumn{2}{|c|}{$O_2$}&\multicolumn{2}{|c|}{$O_1$}&\multicolumn{2}{|c|}{$O_2$}&\multicolumn{2}{|c|}{$O_1$}&\multicolumn{2}{|c|}{$O_2$}\\ 
\hline
CCP&\multicolumn{2}{|c|}{$142$}&\multicolumn{2}{|c|}{$156$}&\multicolumn{2}{|c|}{$60$}&\multicolumn{2}{|c|}{$64$}&
56&8&
61&6\\ 
\hline
Percent &0&898&0&987&0&895&0&955&0&847&0&919\\
\hline
\end{tabular}
\caption{Correctly classified patterns (CCP) of the ANN for the signals generated with parameters $w=0.5,w'=0.25$ and varying $a$ between [$0.5,1$] and $E$ between [$0,1$]. We present the results for 158 training patterns, 67 validation patterns and the average of 5 training sets of 67 patterns each one.}
\label{2N}
\end{center}
\end{table}

\section{Final remarks}
In this article, we have carried out an analysis of BO in the linear, monoatomic chain at the nearest and next-to-nearest neighbors level within a tight-binding approach. We have  developed an efficient machine learning  procedure using ANN to explore the IBO problem. We develop a training process of the ANN such that, according to our error functions, the accuracy of predictions is optimized for the ANN internal parameters in a validation stage. Then the predictability of the ANN is tested directly from {\em ``experimental''} curves derived from directly integrating the equations of motion~(\ref{1v}) and~(\ref{2v}) for given parameters of the electric field strength $E$ and interatomic distance $a$. Retaining $a$ fixed, we have been able to predict the electric field strength of a random trajectory sample with 91\% of accuracy for the 1N level. Similar accuracy of predictions was obtained for fixed $E$ but varying interatomic distance $a$. When both the parameters were allowed to change, at the 2N level, our predictions are accurate up to 84\% for the interatomic distance and 93\% for the electric field strength. Because of the performance of the ANN, these findings are encouraging to extend a similar strategy to explore IBO in realistic systems. Graphene~\cite{BOg1,BOg2} and topological insulators~\cite{BOtop} characterized by linear dispersion relations at low energies are natural candidates for this purpose. Initial work in this direction is being carried out and results will be presented elsewhere. 

\begin{acknowledgements}
We acknowledge support from CONACyT grant 256494 and CIC-UMSNH (M\'exico) under grants 4.22 and 4.23. AR acknowledges Y. Arredondo for valuable discussions.
\end{acknowledgements}


\end{document}